\documentclass[runningheads,a4paper]{llncs}

\usepackage{amssymb}
\usepackage{graphicx}

\usepackage{url}
\urldef{\mailsa}\path|{alfred.hofmann, ursula.barth, ingrid.haas, frank.holzwarth,|
\urldef{\mailsb}\path|anna.kramer, leonie.kunz, christine.reiss, nicole.sator,|
\urldef{\mailsc}\path|erika.siebert-cole, peter.strasser, lncs}@springer.com|
\newcommand{\keywords}[1]{\par\addvspace\baselineskip
\noindent\keywordname\enspace\ignorespaces#1}
\begin{document}

\mainmatter  

\title{A reduction from LWE problem to dihedral coset problem}

\titlerunning{}

%
%
\author{Li Fada
\and Bao Wansu
\thanks{2010thzz@sina.com}
\and Fu Xiangqun\and Zhang Yuchao\and Li Tan}
\authorrunning{}

\institute{Information Science and Technology Institute\\Zhengzhou, China}

%
%

\toctitle{}
\tocauthor{}
\maketitle

\begin{abstract}
Learning with Errors (LWE) problems are the foundations for numerous applications in lattice-based cryptography and are provably as hard as approximate lattice problems in the worst case. Here we present a reduction from LWE problem to dihedral coset problem(DCP). We present a quantum algorithm to generate the input of the two point problem which hides the solution of LWE. We then give a new reduction from two point problem to dihedral coset problem on ${D_{{{({n^{13}})}^{n\log n}}}}$. Our reduction implicate that any algorithm solves DCP in subexponential time would lead a quantum algorithm for LWE.
\end{abstract}
\section{Introduction}
Large quantum computer will be a great challenge to computationally secure cryptography, including breaking public-key cryptography such as RSA and ECC, speeding up brute force searching [1] and finding collisions and claws [2], among numerous quantum algorithmic speed-ups [3][4]. To cope with these threats, some cryptosystems [5] are being researched intensely to replace those broken by quantum computers.
\\\indent One such system bases its security on the hardness of certain lattice problems. Since the late 1990＊s, there has been a fast development of the lattice-based cryptography, resulting in many schemes for encryption and digital signature. In recent years, the Learning with Errors (LWE) problem [6] is noticeable, it is defined as follows: fix a size parameter $n \ge 1$, a modulus $q$, typically taken to be polynomial in $n$, and an error probability distribution $\chi $ on ${Z_q}$. Let ${A_{s,\chi }}$ on $\mathbb{Z}_q^n \times {\mathbb{Z}_q}$ be the probability distribution obtained by choosing a vector ${\mathbf{a}} \in \mathbb{Z}_q^n$ uniformly at random, choosing $e \in {\mathbb{Z}_q}$ according to $\chi $, and outputting $({\mathbf{a}},\left\langle {{\mathbf{a}},{\mathbf{s}}} \right\rangle  + e)$, where additions are performed in ${\mathbb{Z}_q}$, we say that an algorithm solves LWE with modulus $q$ and error distribution $\chi $, given an arbitrary number of independent samples from ${A_{s,\chi }}$ it outputs ${\mathbf{s}}$(with high probability). This problem has proved to be a remarkably flexible basis for cryptographic constructions. For example, the public-key encryption schemes secure under chosen-plaintext attacks [7,8], and chosen-ciphertext attacks [9,10], identity-based encryption (IBE) schemes [11, 12]. The LWE problem was also used to show hardness results in learning theory [13]. Reasons for the popularity of LWE include its simplicity as well as convincing theoretical arguments regarding its hardness, namely, it is provably as hard as certain lattice problems in the worst case. In addition, LWE is attractive as it typically leads to efficient implementations, involving low complexity operations. However, no efficient algorithm for LWE problem has been designed so far, therefore the algorithmic improvement on LWE problem is crucial for LWE-based schemes.
\\ \textbf{Previous Work}
\\\indent Several papers contain studies of the algorithms that solve LWE problem. One simple way to solve LWE is to get the approximate formula ${s_1} \approx ...$ (i.e. a pair $({\mathbf{a}},b)$ where ${\mathbf{a}} = (1,0,...,0)$) by using Gaussian elimination, then the value of ${s_1}$ can be recovered. Iterate this procedure until all ${s_i}$ are recovered. This algorithm require ${2^{O(n\log n)}}$ equations, and with a similar running time. Assume the error distribution is normal, the maximum likelihood algorithm that proposed to solve LPN problem can find the correct $s$ that approximately satisfies the equations after about $O(n)$ equations. This algorithm runs in time ${2^{O(n\log n)}}$. Blum, Kalai, and Wasserman [14] give an algorithm that requires only ${2^{O(n)}}$ samples and time. It allows finding a small set  of equations among ${2^{O(n)}}$ equations, such that $\sum\nolimits_S {{a_i}} $ is, say$(1,0,...,0)$, sum these equations and the first coordinate of $s$ can be recovered. Arora and Ge [15] proposed an algebraic technique for solving LWE, with total complexity (time and space) of ${2^{O({\sigma ^2})}}$, Regev proved [6] that when $\sigma  \ge \sqrt n $ the LWE problem is as hard as the worst-case hardness of standard lattice problems such as GapSVP, thus it remains exponential. There is no subexponential-time algorithm for LWE problem due to the fact that the best known algorithms for lattice problems [16,17] require ${2^{O(n)}}$ time.
\\ \textbf{Our Contribution}
\\\indent Since LWE can be reduced to some hard lattice problems, there was no known efficient (quantum) algorithm. Our main contribution is a reduction from LWE problem to dihedral coset problem. A beautiful reduction from $O({n^{2.5}}) - uSVP$ to the DCP was presented previously by Regev [18]. The reduction uses the property of the unique shortest vector. This inspires us to search for the property of the solution to LWE problem. Then we use the property to design a quantum algorithm, it can create one register a quantum state that hides a fixed information about the solution, then iterate this algorithm $poly(n)$ times to get enough quantum states above as the input of two point problem. Solving this problem needs a reduction to DCP. However, this method always reduce the LWE problem to the DCP on a large dihedral group , even a subexponential algorithm for DCP would not lead an efficient algorithm. Hence we give a new reduction from the two point problem to a dihedral coset problem on ${D_{{{({n^{13}})}^{n\log n}}}}$. Any improvement on the algorithm for DCP will makes a notable impact on LWE problem.
\\ \textbf{Outline}
\\\indent This paper is organized as follows: Section 2 is the preliminaries where some notations and useful lemmas are included. In Section 3, we reduce the LWE problem to find a vector on $q - ary$ lattice. Furthermore, the property of this vector is presented. In Section 4, the reduction from LWE problem to two point problem is described. Section 5, a new reduction from two point problem to dihedral coset problem is given. Section 6 concludes this paper.

\section{Preliminaries}
We denote the notation ${\mathbf{a}}$ as a vector, $a$ as a real number, $\left\| {\mathbf{\sigma }} \right\|$ as the ${l_2}$ norm of ${\mathbf{\sigma }}$, $dist({\mathbf{y}},L({\mathbf{B}}))$ as the distance of $y$ to the lattice $L({\mathbf{B}})$. ${\lambda _1}(B)$ as the length of the shortest vector of a lattice $L({\mathbf{B}})$. Occasionally, we omit the normalization of quantum states.
\\\indent We give some details about the LWE problem here. In all applications, the error probability distribution $\chi $ of LWE problem is chosen to be a discrete Gaussian distribution with mean 0 and standard deviation $\sigma  = \alpha q$ for some $\alpha $, as the LWE problem is hard when $\sigma  \ge \sqrt n $, for simplicity, we consider the case that $\sigma  = \sqrt n $, the case that $\sigma  > \sqrt n $ will also applied. In general, $q$ is an odd prime and taken to be polynomial in $n$, $m$ is the number of samples from ${A_{s,\chi }}$, it is insignificant as the hardness of the problem is independent of it.
\\\indent Our algorithm is related closely to the dihedral coset problem, thus we present the corresponding problems as follows.
\\ \textbf{Definition 1}(Dihedral Coset Problem)
\\\indent Let $d \in {\mathbb{Z}_N}$.The dihedral coset problem (DCP) is to find the value of $d$ given a black box that outputs polynomial states $\frac{1}{{\sqrt 2 }}(\left| x \right\rangle \left| 0 \right\rangle  + \left| {x + d} \right\rangle \left| 1 \right\rangle )$ for a random $x \in {\mathbb{Z}_N}$.
\\\indent We note that Kuperberg [19] and Regev[18] presented the algorithms which solved hidden subgroup problem by sampling cosets on ${D_N}$. However,their algorithm needs $2^{O(\sqrt n)}$ states above which is hard for generating, hence they could not work for solving DCP.
\\ \textbf{Definition 2}[20] (Two Point Problem)
    \indent \\The input to the two point problem consists of $n\log M$ registers. Each register in the state $\frac{1}{{\sqrt 2 }}(\left| 0 \right\rangle \left| {\mathbf{a}} \right\rangle  + \left| 1 \right\rangle \left| {{\mathbf{a}}'} \right\rangle )$ on $1 + n\left\lceil {\log M} \right\rceil $ qubits where ${\mathbf{a}},{\mathbf{a}}' \in {\{ 0,...,M - 1\} ^n}$ are arbitrary such that ${\mathbf{a}} - {\mathbf{a}}'$ is fixed. We say that an algorithm solves the two point problem if it outputs ${\mathbf{a}} - {\mathbf{a}}'$ with $poly(\frac{1}{{n\log M}})$ and time $poly(n\log M)$.
\\ \textbf{Lemma 1}[20] If an algorithm that solves the DCP on dihedral group ${D_{{2^{(n + 1)n}}}}$ exists then there is an algorithm that solves the two point problem with $M = {2^n}$.
\\\indent Regev give this definition and lemma with a failure parameter to make their results more general, as it is insignificant for our algorithm, we omit this parameter when we use the lemma.
\section{Property of the Target Lattice Vector}
As LWE problem is an instance of BDD problem, we can reduce the LWE problem to find a vector on $q - ary$ lattice, and give a useful property of this vector which we can use to design the oracle in the followed quantum algorithm.
\\\indent The LWE problem can also be stated as follows, given $({\mathbf{A}},{\mathbf{v}} = {\mathbf{As}} + {\mathbf{e}}\bmod q)$ where ${\mathbf{A}} \in \mathbb{Z}_q^{m \times n}$, ${\mathbf{s}} \in \mathbb{Z}_q^n$ are chosen randomly, ${\mathbf{e}} \in \mathbb{Z}_q^m$ is chosen based on error probability distribution, the task is to recover ${\mathbf{s}}$. Wang Xiaoyun give a ${\lambda _2} - gap$ estimation of LWE lattices by using the embedding technique, we will also solve the LWE problem on the LWE-based lattice.
\\\indent The $q - ary$ lattice is defined as ${\Lambda _q}({{\mathbf{A}}^T}) = \{ {\mathbf{y}} \in {\mathbb{Z}^m}:{\mathbf{y}} = {\mathbf{As}}\bmod q\begin{array}{*{20}{c}}
  {}
\end{array}for\begin{array}{*{20}{c}}
  {}
\end{array}{\mathbf{s}} \in \mathbb{Z}_q^n\} $ with LLL reduced basis ${\mathbf{B}} = ({{\mathbf{b}}_{\mathbf{1}}}{\mathbf{,}}...{\mathbf{,}}{{\mathbf{b}}_{\mathbf{m}}})$, let ${\mathbf{v}} = \sum\limits_{i = 1}^m {{v_i}{{\mathbf{e}}_i}} $ be the representation in the orthonormal basis of ${\mathbf{v}} = {\mathbf{As}} + {\mathbf{e}}\bmod q$ where $({{\mathbf{e}}_{\mathbf{1}}}{\mathbf{,}}...{\mathbf{,}}{{\mathbf{e}}_{\mathbf{m}}})$ is the orthonormal basis. The LWE problem can then be restated as: given ${\mathbf{v}}$ which is the sum of a lattice point ${\mathbf{u}} = \sum\limits_{i = 1}^m {{\alpha _i}{{\mathbf{b}}_i}} $ and a short ※noise vector§ ${\mathbf{e}}$, find the ※closest§ lattice vector ${\mathbf{u}}$. We use the embedding technique to construct the $m + 1$ rank lattice $L({\mathbf{B}}') = L(({\mathbf{b}}{{\mathbf{'}}_{\mathbf{1}}}{\mathbf{,}}...{\mathbf{b}}{{\mathbf{'}}_{\mathbf{m}}}{\mathbf{,b}}{{\mathbf{'}}_{{\mathbf{m + 1}}}}))$ which called LWE-based lattice, the basis $({\mathbf{b}}{{\mathbf{'}}_{\mathbf{1}}}{\mathbf{,}}...{\mathbf{b}}{{\mathbf{'}}_{\mathbf{m}}}{\mathbf{,b}}{{\mathbf{'}}_{{\mathbf{m + 1}}}})$ can be represented as $\left( {\begin{array}{*{20}{c}}
  {\mathbf{B}}&{\mathbf{v}} \\
  {{0^T}}&\eta
\end{array}} \right)$ where $\eta $ is an indeterminate parameter. For any ${\mathbf{x}}' \in L({\mathbf{B}}')$, there exists ${\mathbf{x}} = \sum\limits_i {{x_i}{{\mathbf{b}}_i}}  \in L({\mathbf{B}})$ such that ${\mathbf{x}}' = ({\mathbf{x}} - t{\mathbf{v}}, - t\eta )$, especially, let ${\mathbf{u}} = \sum\limits_{i = 1}^m {{\alpha _i}{{\mathbf{b}}_i}} $, then the vector $({\mathbf{v}} - {\mathbf{u}},\eta ) = ({\mathbf{b}}{{\mathbf{'}}_{\mathbf{1}}}{\mathbf{,}}...{\mathbf{b}}{{\mathbf{'}}_{\mathbf{m}}}{\mathbf{,b}}{'_{{\mathbf{m + 1}}}})\left( {\begin{array}{*{20}{c}}
  { - {\alpha _1}} \\
   \vdots  \\
  { - {\alpha _m}} \\
  1
\end{array}} \right) \in L({\mathbf{B}}')$. Notice that if we can find this lattice vector, we will solve the LWE problem with high probability.
\\\indent As there exists ${\mathbf{v}} - {\mathbf{u}} = {\mathbf{e}}$, the main hardness of solving ${\mathbf{v}} = {\mathbf{As}} + {\mathbf{e}}\bmod q$ is the uncertainty of the ＆error＊. Correspondingly we first research the property of the error vector ${\mathbf{e}}$ so that we can get some useful properties of the target lattice vector $({\mathbf{v}} - {\mathbf{u}},\eta )$. The error vector ${\mathbf{e}}$ is chosen from a discrete Gaussian distribution, therefore the tail bound for discrete Gaussian distribution in the following lemma will be useful in estimating the norm of the error vector.
\\ \textbf{Lemma 2}[21] Let $d > 1,s > 0$ and $n$ be a positive integer. Let ${\mathbf{\sigma }} \in {\mathbb{Z}^n}$ be randomly chosen according to ${D_{{\mathbb{Z}^n},s}}$. Then there exists $\Pr [\left\| {\mathbf{\sigma }} \right\| \ge d\frac{{s\sqrt n }}{{\sqrt {2\pi } }}] \le {(d \cdot \exp (\frac{{1 - {d^2}}}{2}))^n}$.
\\\indent By the lemma 2,as ${\mathbf{e}}$ is randomly chosen from ${D_{{\mathbb{Z}^m},\alpha q}}$, set $d = 1 + \frac{\varepsilon }{2}$ where $\frac{1}{{n\sqrt {1 + \frac{1}{{2n}}}  + n - \frac{1}{2}}} < \varepsilon  < 1$, then we get $\left\| {\mathbf{e}} \right\| \le (1 + \frac{\varepsilon }{2})\alpha q\sqrt {\frac{m}{{2\pi }}} $ with high probability.
\\\indent As the target lattice vector $({\mathbf{v}} - {\mathbf{u}},\eta )$ has an indeterminate parameter $\eta $ which influence the norm of $({\mathbf{v}} - {\mathbf{u}},\eta )$, it is also important to choose an appropriate one. Next lemma presents the selection of this parameter.
\\ \textbf{Lemma 3}[22] There exists a polynomial-time algorithm that, given $\forall {\mathbf{y}} \in {R^m}$ and a lattice $L({\mathbf{B}})$, outputs a lattice vector ${\mathbf{c}} \in L({\mathbf{B}})$ such that $\left\| {{\mathbf{y}} - {\mathbf{c}}} \right\| \in [dist({\mathbf{y}},L({\mathbf{B}})), \le {2^n}dist({\mathbf{y}},L({\mathbf{B}}))]$.
\\\indent For ${\mathbf{v}} = {\mathbf{As + e}}\bmod q$, we can find a vector ${\mathbf{c}} \in {\Lambda _q}({{\mathbf{A}}^T})$ such that $\left\| {\mathbf{e}} \right\| \le d \le {2^n}\left\| {\mathbf{e}} \right\|$ by the lemma 2 where $d = \left\| {{\mathbf{v}} - {\mathbf{c}}} \right\|$. Consider the set $S = \{ \frac{d}{{{{(1 + \frac{1}{n})}^i}}}:0 \le i \le \log _{1 + \frac{1}{n}}^{{2^n}}\} $ which is  polynomial-sized, it can be proved that there exists ${i_0}$ such that $(1 - \frac{1}{n})\left\| {\mathbf{e}} \right\| \le \frac{d}{{{{(1 + \frac{1}{n})}^{{i_0}}}}} \le (1 + \frac{1}{n})\left\| {\mathbf{e}} \right\|$, we choose $\eta  = \frac{d}{{{{(1 + \frac{1}{n})}^{{i_0}}}}}$. Therefore $\left\| {({\mathbf{v}} - {\mathbf{u}},\eta )} \right\| = \sqrt {{{\left\| {\mathbf{e}} \right\|}^2} + {\eta ^2}}  \le \left\| {\mathbf{e}} \right\| + \frac{d}{{{{(1 + \frac{1}{n})}^{{i_0}}}}}$ and we get $\left\| {({\mathbf{v}} - {\mathbf{u}},\eta )} \right\| \le (2 + \frac{1}{n})\left\| {\mathbf{e}} \right\|$.
\\\indent Next lemma gives the lower bound of the shortest vector on $q - ary$ lattice.
\\\textbf{Lemma 4}[23] Let $n,m \in \mathbb{Z}$, and $q$ be a prime such that $m > n,{q^{1 - \frac{{n + c}}{m}}} > \sqrt {\pi {e^{1 + 2\omega }}} $ for some positive constants $c$ and $w < 1.024 \times {10^{ - 4}}$. Let ${\mathbf{A}} \in \mathbb{Z}_q^{m \times n}$ be chosen uniformly. Then for any ${\mathbf{x}} \in {R^m}$ we have, with probability bigger than $1 - {q^{ - c}}$,$\mathop {\min }\limits_{{\mathbf{a}} \in {\Lambda _q}({{\mathbf{A}}^T})} \left\| {{\mathbf{a}} - {\mathbf{x}}} \right\| \ge T$. In particular, ${\lambda _1}({\Lambda _q}({{\mathbf{A}}^T})) \ge T$.
\\\indent Now we are ready to give the property of $({\mathbf{v}} - {\mathbf{u}},\eta )$.
\\ \textbf{Theorem 1} For any ${\mathbf{x}}' = ({\mathbf{x}} - t{\mathbf{v}}, - t\eta ) \in L({\mathbf{B}}')$, if the vector is not a multiple of $({\mathbf{v}} - {\mathbf{u}},\eta )$, then $\left\| {{\mathbf{x}}'} \right\| > \frac{T}{{\sqrt 2 }}$.
\\\textbf{Proof:} we will use proof of contradiction. If the vector ${\mathbf{x}}' = ({\mathbf{x}} - t{\mathbf{v}}, - t\eta )$ is not a multiple of $({\mathbf{v}} - {\mathbf{u}},\eta )$ and $\frac{T}{{\sqrt 2 }} > \left\| {{\mathbf{x}}'} \right\| = \sqrt {{{\left\| {{\mathbf{x}} - t{\mathbf{v}}} \right\|}^2} + {{(t\eta )}^2}} $, then we have $\left\| {{\mathbf{x}} - t{\mathbf{v}}} \right\| < \sqrt {\frac{{{T^2}}}{2} - {{(t\eta )}^2}} $, as ${\mathbf{x}}'$ is not parallel to $({\mathbf{v}} - {\mathbf{u}},\eta )$, ${\mathbf{x}} - t{\mathbf{u}} \in {\Lambda _q}({{\mathbf{A}}^T})$ is a non-zero lattice vector, and we have the following inequality
$$\left\| {{\mathbf{x}} - t{\mathbf{u}}} \right\| \le \left\| {{\mathbf{x}} - t{\mathbf{v}}} \right\| + t\left\| {{\mathbf{u}} - {\mathbf{v}}} \right\| < \sqrt {\frac{{{T^2}}}{2} - {{(t\eta )}^2}}  + t\left\| {\mathbf{e}} \right\|$$
\\\indent The last term of the above inequality is maximized when $t = \frac{T}{{\left\| {\mathbf{e}} \right\|\sqrt {2(1 + \frac{1}{n})(2 + \frac{1}{n})} }}$, and therefore for all $t$, we have $\left\| {{\mathbf{x}} - t{\mathbf{u}}} \right\| < \frac{T}{2}(\sqrt {\frac{{2n + 1}}{{2n + 2}}}  + \sqrt {\frac{{2{n^2}}}{{(n + 1)(2n + 1)}}} ) < T$, by the lemma 4, we have ${\lambda _1}({\Lambda _q}({{\mathbf{A}}^T})) \ge T$, then it gives us the contradiction that $\left\| {{\mathbf{x}} - t{\mathbf{u}}} \right\| < {\lambda _1}({\Lambda _q}({{\mathbf{A}}^T}))$.
\section{The Quantum Algorithm for LWE Problem}
In section 3, solving LWE problem, that is, getting ${\mathbf{u}} = \sum\limits_{i = 1}^m {{\alpha _i}{{\mathbf{b}}_i}} $, relies on the property of a special vector $({\mathbf{v}} - {\mathbf{u}},\eta )$ on the $q - ary$ lattice. In this section we reduce the problem to the two point problem based on this property of $({\mathbf{v}} - {\mathbf{u}},\eta )$. More specifically, we start by creating a superposition of many lattice points, and then we design an oracle so that state collapses to a superposition $\frac{1}{{\sqrt 2 }}(\left| {0,{\mathbf{a}}} \right\rangle  + \left| {1,{\mathbf{a}}'} \right\rangle )$ by the measurement. This two vectors ${\mathbf{a}}$ and ${\mathbf{a}}'$ hides a fixed difference that ${a_i} - {a_i}' = {\alpha _i}\begin{array}{*{20}{c}}  {}&{i \in \{ 1,...,m\} }
\end{array}$ where ${a_i},{a_i}'$ are the $i'th$ entry of ${\mathbf{a}},{\mathbf{a}}'$ in the basis $({\mathbf{b}}{{\mathbf{'}}_{\mathbf{1}}}{\mathbf{,}}...{\mathbf{b}}{{\mathbf{'}}_{\mathbf{m}}}{\mathbf{,b}}{{\mathbf{'}}_{{\mathbf{m + 1}}}})$, we can not obtain the information about the $(m + 1)'th$ entry of ${\mathbf{a}} - {\mathbf{a}}'$, however it provides no information about the solution of LWE, hence the last $\left\lceil {\log M} \right\rceil $ qubits are discarded, then this register is in the state of $\frac{1}{{\sqrt 2 }}(\left| {0,{a_1},...,{a_m}} \right\rangle  + \left| {1,{a_1}',...,{a_m}'} \right\rangle )$. Repeating this procedure $O(m\log M)$ times creates a complete input to the two point problem whose solution is$({\alpha _1},...,{\alpha _m})$.
\\\indent Let $\forall {\mathbf{a}} \in {\{ 0,1,...M - 1\} ^{m + 1}}$ where $M = {2^n}$, $t \in \{ 0,1\} $, ${w_i}$ be $m + 1$ real values in $[0,1)$. In the following, we will give a quantum algorithm that makes one register in the input to the two point problem that hides the difference $ \pm ({\alpha _1},...,{\alpha _m})$.
\\\indent First, we use the Hardmard transform to get a superposition $$\frac{1}{{\sqrt {2{M^{m + 1}}} }}\sum\limits_{t \in \{ 0,1\} ,{\mathbf{a}} \in {{\{ 0,...,M - 1\} }^{m + 1}}}^{} {\left| {t,{\mathbf{a}}} \right\rangle } $$ \\note that ${\mathbf{a}} = ({a_1},...,{a_{m + 1}}) \in {R^{m + 1}}$ is not a lattice point, correspondingly we define a function $f(t,{\mathbf{a}}) = \sum\limits_{i = 1}^m {{a_i}{\mathbf{b}}{'_i}}  + t{\mathbf{b}}{'_{m + 1}}$ such that $(t,{\mathbf{a}})$ can be related to a lattice point of the lattice $L({\mathbf{B}}')$. Then we should design another function $g({\mathbf{v}})$ where ${\mathbf{v}} \in L(B')$ such that for any output of $g({\mathbf{v}})$, the corresponding inputs will be $(t,{\mathbf{a}}),(t',{\mathbf{a}}')$ where ${\mathbf{a}} - {\mathbf{a'}} = ({\alpha _1},...,{\alpha _m})$ with high probability. Therefore for any lattice point of the lattice $L({\mathbf{B}}')$, considering their representation in the orthonormal basis ${\mathbf{v}} = \sum\limits_{i = 1}^{m + 1} {{v_i}{{\mathbf{e}}_i}} $, design the function $g({\mathbf{v}}) = (\left\lfloor {\frac{{\sqrt {2(m + 1)} {v_1}}}{T} - {w_1}} \right\rfloor ,\left\lfloor {\frac{{\sqrt {2(m + 1)} {v_2}}}{T} - {w_2}} \right\rfloor ,...,\left\lfloor {\frac{{\sqrt {2(m + 1)} {v_{m + 1}}}}{T} - {w_{m + 1}}} \right\rfloor )$, let  $F = g \circ f(t,{\mathbf{a}})$ be the oracle function of the algorithm.
\\\indent Now the quantum algorithm that makes one register in the input to the two point problem is given as follows
\\ \textbf{Step1}  Choosing ${w_i}$ uniformly from $[0,1)$, perform the Hadamard transform on the $1 + (m + 1)\left\lceil {\log M} \right\rceil $ qubits data register to get the equal superposition $$\left| {{\varphi _1}} \right\rangle  = \frac{1}{{\sqrt {2{M^{m + 1}}} }}\sum\limits_{t \in \{ 0,1\} , {\mathbf{a}} \in {{\{ 0,...,M - 1\} }^{m + 1}}}^{} {\left| {t,{\mathbf{a}}} \right\rangle } $$
\\ \textbf{Step2}  Give one $A = (m + 1)\left\lceil {\log M} \right\rceil $ qubits target register initialized to $\left| 0 \right\rangle $ and apply the black box to performs the operation $U\left| {t,{\mathbf{a}}} \right\rangle \left| 0 \right\rangle  = \left| {t,{\mathbf{a}}} \right\rangle \left| {0 \oplus F(t,{\mathbf{a}})} \right\rangle $
\\ \textbf{Step3}  Measure the target register and assumes we get the results $({r_1},{r_2},...,{r_{m + 1}})$.
\\In the following two theorems, we will prove that after the measurement, the states in the data register will collapse to $\left| {{\varphi _2}} \right\rangle  = \frac{1}{{\sqrt 2 }}(\left| {0,{\mathbf{a}}} \right\rangle  + \left| {1,{\mathbf{a}}'} \right\rangle )$ where ${a_i} - {a_i}' =  \pm {\alpha _i}\begin{array}{*{20}{c}}
  {}&{i \in \{ 1,...,m\} }
\end{array}$ with high probability.
\\ \textbf{Theorem 2} For any result $({r_1},{r_2},...,{r_{m + 1}})$ from the measurement, the $(t,{\mathbf{a}})$ that satisfies $F = g \circ f(t,{\mathbf{a}}) = ({r_1},{r_2},...,{r_{m + 1}})$ has only three conditions: $\left| {0,{\mathbf{a}}} \right\rangle $, $\left| {1,{\mathbf{a}}'} \right\rangle $ or $\frac{1}{{\sqrt 2 }}(\left| {0,{\mathbf{a}}} \right\rangle  + \left| {1,{\mathbf{a}}'} \right\rangle )$ where ${a_i} - {a_i}' =  \pm {\alpha _i}$ㄛ$i \in \{ 1,...,m\} $.
\\ \textbf{Proof:} for $\forall (t,{\bf{a}}),(t',{\bf{a}}')$, $ \mathrel\backepsilon  g \circ f(t,{\bf{a}}) = g \circ f(t',{\bf{a}}') = ({r_1},{r_2},...,{r_{m + 1}})$, the representation in the orthonormal basis of $f(t,{\mathbf{a}})$ and $f(t',{\mathbf{a}}')$ are ${\mathbf{c}} = \sum\limits_{i = 1}^{m + 1} {{c_i}{{\mathbf{e}}_i}} $, ${\mathbf{c}}' = \sum\limits_{i = 1}^{m + 1} {{c_i}'{{\mathbf{e}}_i}} $ respectively. If ${\mathbf{c}} - {\mathbf{c}}'$ is a non-parallel lattice point to $({\mathbf{v}} - {\mathbf{u}},\eta )$, then $\left\| {{\mathbf{c}} - {\mathbf{c}}'} \right\| > \frac{T}{{\sqrt 2 }}$ by the theorem 1, that is, $\sqrt {{{({c_1} - {c_1}')}^2} + ... + {{({c_{m + 1}} - {c_{m + 1}}')}^2}}  > \frac{T}{{\sqrt 2 }}$. There exist a coordinate $i$ such that $\sqrt {(m + 1)({{({c_i} - {c_i}')}^2})}  > \frac{T}{{\sqrt 2 }}$ and we have $\left| {{c_i} - {c_i}'} \right| > \frac{T}{{\sqrt {2(m + 1)} }}$, assume without loss of generality that ${c_i} > {c_i}'$, therefore ${c_i} > {c_i}' + \frac{T}{{\sqrt {2(m + 1)} }}$, considering $g({\mathbf{c}})$ and $g({\mathbf{c}}')$, because $\frac{{\sqrt {2(m + 1)} {c_i}}}{T} > \frac{{\sqrt {2(m + 1)} [{c_i}' + \frac{T}{{\sqrt {2(m + 1)} }}]}}{T}$ which implies $\frac{{\sqrt {2(m + 1)} {c_i}}}{T} > \frac{{\sqrt {2(m + 1)} {c_i}'}}{T} + 1$, for any randomly chose ${w_i}$, there will be $\left\lfloor {\frac{{\sqrt {2(m + 1)} {c_i}}}{T} - {w_i}} \right\rfloor  \ne \left\lfloor {\frac{{\sqrt {2(m + 1)} {c_i}'}}{T} - {w_i}} \right\rfloor $ which implies $g({\mathbf{c}}) \ne g({\mathbf{c}}')$, that is $g \circ f(t,{\mathbf{a}}) \ne g \circ f(t',{\mathbf{a}}')$, it gives us the contradiction that $g \circ f(t,{\mathbf{a}}) = g \circ f(t',{\mathbf{a}}') = ({r_1},{r_2},...,{r_{m + 1}})$. Hence ${\mathbf{c}} - {\mathbf{c}}'$ is parallel to $({\mathbf{v}} - {\mathbf{u}},\eta )$, set ${\mathbf{c}} - {\mathbf{c}}' = k({\mathbf{v}} - {\mathbf{u}},\eta )$ for some integer $k \ne 0$. Considering the lattice point with the representation in the basis $({\mathbf{b}}{{\mathbf{'}}_{\mathbf{1}}}{\mathbf{,}}...{\mathbf{b}}{{\mathbf{'}}_{\mathbf{m}}}{\mathbf{,b}}{{\mathbf{'}}_{{\mathbf{m + 1}}}})$, we get the following equation
$$\sum\limits_{i = 1}^m {{a_i}{\mathbf{b}}{'_i}}  + t{\mathbf{b}}{'_{m + 1}} - (\sum\limits_{i = 1}^m {{a_i}'{\mathbf{b}}{'_i}}  + t'{\mathbf{b}}{'_{m + 1}}) = k(\sum\limits_{i = 1}^m { - {\alpha _i}{\mathbf{b}}{'_i}}  + {\mathbf{b}}{'_{m + 1}})$$
\\By considering the coordinate of ${\mathbf{b}}{'_{m + 1}}$, we will obtain $t - t' = k$. Correspondingly, if $t = t'$, there exists $\left| {t,{\mathbf{a}}} \right\rangle $ and $\left| {t,{\mathbf{a}}'} \right\rangle $ where $t = 0,1$ such that $g \circ f(t,{\mathbf{a}}) = g \circ f(t,{\mathbf{a}}') = ({r_1},{r_2},...,{r_{m + 1}})$, as $k = 0$, there exists ${\mathbf{c}} - {\mathbf{c}}' = 0$ which implies ${\mathbf{a}} = {\mathbf{a'}}$. If $t \ne t'$, then $t - t' = 1$, there exists $\sum\limits_{i = 1}^m {({a_i} - {a_i}'){{\mathbf{b}}_i}'}  - {{\mathbf{b}}_{m + 1}}' = \sum\limits_{i = 1}^m {{\alpha _i}{\mathbf{b}}{'_i}}  - {\mathbf{b}}{'_{m + 1}})$. Therefore for $i \in \{ 1,...,m\} $, there exists ${a_i} - {a_i}' = {\alpha _i}$.                              \\\indent For $\left| {t,{\mathbf{a}}} \right\rangle $ that satisfies $F = g \circ f(t,{\mathbf{a}}) = ({r_1},{r_2},...,{r_{m + 1}})$, only the condition $\frac{1}{{\sqrt 2 }}(\left| {0,{\mathbf{a}}} \right\rangle  + \left| {1,{\mathbf{a}}'} \right\rangle )$ that includes the information about the solution of LWE problem. Hence we need to consider the problem that for any $\bar r = ({r_1},{r_2},...,{r_{m + 1}})$, whether the possibility of $F(1 - t,{\mathbf{a}}) = F(t,{\mathbf{a}}') = \bar r$ is acceptable.
\\ \textbf{Theorem 3} With probability at least $1 - A\frac{{\log n\sqrt {\log n} }}{{{n^\delta }}} - \frac{{6\alpha qm}}{{\sqrt {2\pi } {2^n}}}$ where $\delta  \ge 3$, $A = \frac{{2(2 + \frac{1}{n})(1 + \frac{\varepsilon }{2})}}{{\sqrt \pi  }}$, there exists $(1 - t,{\mathbf{a}})$ and $(t,{\mathbf{a}}')$ such that $F(1 - t,{\mathbf{a}}) = F(t,{\mathbf{a}}') = \bar r$.
\\ \textbf{Proof:} We assume $t = 0$, the proof for $t = 1$ is similar. If there exists $(0,{\mathbf{a}})$ that satisfies $g \circ f(0,{\mathbf{a}}) = ({r_1},{r_2},...,{r_{m + 1}})$ where ${\mathbf{a}} = ({a_1},...,{a_m},{a_{m + 1}})$, we consider the probability $P$ that $g \circ f(1,{\mathbf{a}}') = ({r_1},{r_2},...,{r_{m + 1}})$ where ${\mathbf{a}}' = ({a_1} - {\alpha _1},...,{a_m} - {\alpha _m},{a_{m + 1}}')$, ${a_{m + 1}}'$ is arbitrary. First, ${\mathbf{a}}' = ({a_1} - {\alpha _1},...,{a_m} - {\alpha _m},{a_{m + 1}}')$ should be an element of ${\{ 0,1,...M - 1\} ^{m + 1}}$. According to Lemma 3, $\left\| {({\mathbf{v}} - {\mathbf{u}},\eta )} \right\| \le (2 + \frac{1}{n})\left\| {\mathbf{e}} \right\|$, then we have $\sqrt {m{\alpha _i}^2 + 1}  \le (2 + \frac{1}{n})\left\| {\mathbf{e}} \right\| \le (2 + \frac{1}{n})(1 + \frac{\varepsilon }{2})\alpha q\sqrt {\frac{m}{{2\pi }}}  \le 6\alpha q\sqrt {\frac{m}{{2\pi }}} $, correspondingly ${\alpha _i} \le \frac{{6\alpha q}}{{\sqrt {2\pi } }}$. The possibility that ${a_i} - {\alpha _i} \notin \{ 0,1,...M - 1\} $ is $\frac{{6\alpha q}}{{\sqrt {2\pi } M}}$, similarly for $i \in \{ 1,...,m\} $, the possibility that ${\mathbf{a}}' = ({a_1} - {\alpha _1},...,{a_m} - {\alpha _m},{a_{m + 1}}') \notin {\{ 0,1,...M - 1\} ^{m + 1}}$ is $\sum\limits_{i = 1}^m {\frac{{6\alpha q}}{{\sqrt {2\pi } M}}}  = \frac{{6\alpha qm}}{{\sqrt {2\pi } M}}$.
\\\indent Then consider the possibility that $g \circ f(0,{\mathbf{a}}) \ne g \circ f(1,{\mathbf{a}}')$. Notice that $f(0,{\mathbf{a}}) - f(1,{\mathbf{a}}') = ({\mathbf{v}} - {\mathbf{u}},\eta )$, we assume the possibility that $g \circ f(0,{\mathbf{a}})$ and $g \circ f(1,{\mathbf{a}}')$ differ on the i'th coordinate is ${P_i}$,  the i'th coordinate of $g \circ f(0,{\mathbf{a}})$ and $g \circ f(1,{\mathbf{a}}')$ are $\frac{{\left\langle {f(0,{\mathbf{a}}),{{\mathbf{e}}_i}} \right\rangle  \cdot \sqrt {2(m + 1)} }}{T}$ and $\frac{{\left\langle {f(1,{\mathbf{a}}'),{{\mathbf{e}}_i}} \right\rangle  \cdot \sqrt {2(m + 1)} }}{T}$ respectively. We can find that $\frac{{\left\langle {f(0,{\mathbf{a}}),{{\mathbf{e}}_i}} \right\rangle  \cdot \sqrt {2(m + 1)} }}{T} - \frac{{\left\langle {f(1,{\mathbf{a}}'),{{\mathbf{e}}_i}} \right\rangle  \cdot \sqrt {2(m + 1)} }}{T} = \frac{{\left\langle {({\mathbf{v}} - {\mathbf{u}},\eta ),{{\mathbf{e}}_i}} \right\rangle  \cdot \sqrt {2(m + 1)} }}{T}$. And we set $l = (\frac{{\left\langle {f(0,{\mathbf{a}}),{{\mathbf{e}}_i}} \right\rangle  \cdot \sqrt {2(m + 1)} }}{T} - {w_i}) - \left\lfloor {\frac{{\left\langle {f(0,{\mathbf{a}}),{{\mathbf{e}}_i}} \right\rangle  \cdot \sqrt {2(m + 1)} }}{T} - {w_i}} \right\rfloor $, then the i'th coordinate of $g \circ f(0,{\mathbf{a}})$ and $g \circ f(1,{\mathbf{a}}')$ are the same only when $\frac{{\left\langle {({\mathbf{v}} - {\mathbf{u}},\eta ),{{\mathbf{e}}_i}} \right\rangle  \cdot \sqrt {2(m + 1)} }}{T} + l < 1$, since ${w_i}$ are randomly chosen, ${P_i} = \frac{{\left\langle {({\mathbf{v}} - {\mathbf{u}},\eta ),{{\mathbf{e}}_i}} \right\rangle  \cdot \sqrt {2(m + 1)} }}{T}$.
\\\indent There exists $T = \frac{T}{{(2 + \frac{1}{n})(1 + \frac{\varepsilon }{2})\alpha q\sqrt {\frac{m}{{2\pi }}} }} \cdot (2 + \frac{1}{n})(1 + \frac{\varepsilon }{2})\alpha q\sqrt {\frac{m}{{2\pi }}} $ and $\left\| {({\mathbf{v}} - {\mathbf{u}},\eta )} \right\| \le (2 + \frac{1}{n})(1 + \frac{\varepsilon }{2})\alpha q\sqrt {\frac{m}{{2\pi }}} $, hence
$T \ge \frac{T}{{(2 + \frac{1}{n})(1 + \frac{\varepsilon }{2})\alpha q\sqrt {\frac{m}{{2\pi }}} }} \cdot \left\| {({\mathbf{v}} - {\mathbf{u}},\eta )} \right\|$
\\Correspondingly ${P_i} \le \frac{{\left\langle {({\mathbf{v}} - {\mathbf{u}},\eta ),{{\mathbf{e}}_i}} \right\rangle  \cdot \sqrt {2(m + 1)} (2 + \frac{1}{n})(1 + \frac{\varepsilon }{2})\alpha q\sqrt {\frac{m}{{2\pi }}} }}{{T \cdot \left\| {({\mathbf{v}} - {\mathbf{u}},\eta )} \right\|}}$
\\\indent Therefore the possibility that $g \circ f(0,{\mathbf{a}})$ and $g \circ f(1,{\mathbf{a}}')$ differ is at most
$$P = \sum\limits_{i = 1}^m {{P_i}}  \le \frac{{\sum\limits_{i = 1}^m {\left\langle {({\mathbf{v}} - {\mathbf{u}},\eta ),{{\mathbf{e}}_i}} \right\rangle }  \cdot \sqrt {2(m + 1)} (2 + \frac{1}{n})(1 + \frac{\varepsilon }{2})\alpha q\sqrt {\frac{m}{{2\pi }}} }}{{T \cdot \left\| {({\mathbf{v}} - {\mathbf{u}},\eta )} \right\|}}$$\\\indent For the prime $q \ge O({n^5})$, as $m$ is insignificant for the hardness of LWE problem, we choose $m = n\log n$ which the parameter is chosen in cryptosystem based on LWE, at this time $T = q$. We also use the fact that the ${l_1}$ norm of a vector is at most $\sqrt n $ times its ${l_2}$ norm, therefore we can obtain
$$P \le A\frac{{\log n\sqrt {\log n} }}{{{n^\delta }}}$$
\\where $A = \frac{{2(2 + \frac{1}{n})(1 + \frac{\varepsilon }{2})}}{{\sqrt \pi  }}$ and $\delta  \ge 3$.
\\\indent The sum of two error probabilities is at most $A\frac{{\log n\sqrt {\log n} }}{{{n^\delta }}} + \frac{{6\alpha qm}}{{\sqrt {2\pi } M}}$. Hence the possibility that $F(1 - t,{\mathbf{a}}) = F(t,{\mathbf{a}}') = \bar r$ is ${P_{one}} = 1 - A\frac{{\log n\sqrt {\log n} }}{{{n^\delta }}} - \frac{{6\alpha qm}}{{\sqrt {2\pi } {2^n}}}$.\\\indent As we must iterate this algorithm $m\log M$ times, correspondingly the possibility that we can obtain a complete input to the two point problem is ${P_{com}} = {[1 - A\frac{{\log n\sqrt {\log n} }}{{{n^\delta }}} - \frac{{6\alpha qm}}{{\sqrt {2\pi } M}}]^{m\log M}}$. The figure 1 shows the relationship between ${P_{com}}$ and the size parameter $n$ for different $q$.
\begin{figure}[htbp]
\centering
\includegraphics[width=9.08cm,height=6.8cm]{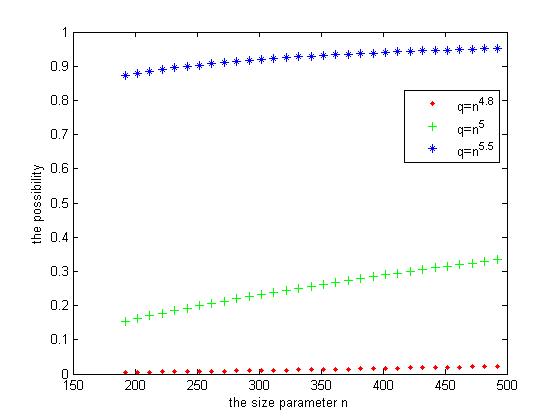}
\caption{relationship between ${P_{com}}$ and the size parameter $n$ for different $q$}
\label{fig1}
\end{figure}
\\\indent We consider the condition of $q = {n^5}$, when $n = 192,233,256,320$. Table 1 shows that for different $n$, the results of ${P_{one}}$ and ${P_{com}}$.
\begin{table}[htbp]
\centering
\caption{the possibility when $q = {n^5}$}
\label{tab:1}       
\begin{tabular}{llll}
\hline\noalign{\smallskip}
n & m & ${P_{one}}$ & ${P_{com}}$  \\
\noalign{\smallskip}\hline\noalign{\smallskip}
192 & 1456 & $1 - 6.68 \times {10^{ - 6}}$ & 0.1545\\
233 & 1832 & $1 - 3.94 \times {10^{ - 6}}$ & 0.1857\\
256 & 2048 & $1 - 3.05 \times {10^{ - 6}}$ & 0.2020\\
320 & 2663 & $1 - 1.66 \times {10^{ - 6}}$ & 0.2438\\
\noalign{\smallskip}\hline
\end{tabular}
\end{table}
\\\indent Now we reduce the LWE problem (with $q \ge O({n^5})$ ) to the two point problem, for simplicity, we only consider the condition of $q = {n^5}$ in the following section. Lemma 1 gives us a conclusion that the two point problem can be reduced to the DCP over a dihedral group ${D_N}$, however, it is always reduced to the problem over a larger ${D_N}$ which leads to an exponential-time algorithm. In the next section, we will improve the reduction from the two point problem whose solution is $({\alpha _1},...,{\alpha _m})$ to DCP over ${D_{{n^{13n\log n}}}}$ which implies that a subexponential time quantum algorithm for DCP will lead a quantum algorithm for LWE with the computation complexity ${2^{O(\sqrt {n\log n\log n} )}}$.
\section{Improved Reduction from Two Point Problem to Dihedral Coset Problem}
The main improvement on the reduction from two point problem to DCP is the mapping $f$. In the previous reduction, given an input to the two point problem, we can create an input to the DCP by using the mapping $f$, that is
$$\frac{1}{{\sqrt 2 }}(\left| 0 \right\rangle \left| {\mathbf{a}} \right\rangle  + \left| 1 \right\rangle \left| {{\mathbf{a}}'} \right\rangle ) \to \frac{1}{{\sqrt 2 }}(\left| 0 \right\rangle \left| {f({\mathbf{a}})} \right\rangle  + \left| 1 \right\rangle \left| {f({\mathbf{a}}')} \right\rangle )$$
\\\indent The mapping $f$ from ${\{ 0,...,M - 1\} ^n}$ to $\{ 0,...,{(2M)^n} - 1\} $ is defined as follows:
$$f({a_1},...,{a_n}) = {a_1} + {a_2} \cdot 2M + ... + {a_n} \cdot {(2M)^{n - 1}}$$
\\\indent As $f$ is a bijective mapping, the difference $f({\mathbf{a}}) - f({\mathbf{a}}')$ obtained by calling the DCP algorithm will give us ${\mathbf{a}} - {\mathbf{a}}'$. However, the range of $f$ is ${(2M)^n}$ so that the dihedral group would be ${D_{{2^{(n + 1)n}}}}$ for $M = {2^n}$. Thus the algorithm will also be exponential even if there exists an algorithm for DCP over the group ${D_N}$ runs in subexponential time ${2^{\sqrt {\log N} }}$.
\\\indent The main reasons for choosing a bijective mapping $f$ are: (1) for every ${\bf{a}} - {\bf{a'}} = ({\alpha _1},...,{\alpha _m})$, ${\mathbf{a}},{\mathbf{a'}} \in {\{ 0,...,M - 1\} ^m}$, a bijective mapping can guarantee $f({\bf{a}}) - f({\bf{a'}}) = f(({\alpha _1},...,{\alpha _m}))$ is a fixed value. (2) We can easily get the vector $({\alpha _1},...,{\alpha _m})$ from $f(({\alpha _1},...,{\alpha _m}))$. (3)it can be implemented as a unitary transform ${U_f}:\left| {\mathbf{a}} \right\rangle  \to \left| {f({\mathbf{a}})} \right\rangle $ which can be regarded as a minimal oracle[24].
\\\indent However, it makes the range of $f$ too large. Thus we construct a mapping that it can meet the three conditions and makes the range of mapping as small as possible.
\\\textbf{Theorem 4} If an algorithm that solves the DCP over dihedral group ${D_{{{({n^{13}})}^{n\log n}}}}$ exists then there is an algorithm that solves the two point problem with ${\mathbf{a}} - {\mathbf{a}}' = ({\alpha _1},...,{\alpha _m})$
\\\textbf{Proof:} First, we give a bijective mapping $g$ for ${\mathbf{b}}$ that belongs to
\\$X = \{{\bf{a}}\left|{{a_i}<{n^{13}},i\in\{ 1,...,m\}}\right.\}$:
$$g({b_1},...,{b_m}) = {b_1} + {b_2} \cdot {n^{13}} + ... + {b_n}{({n^{13}})^{m - 1}}$$
The range of $g$ is $\{ 0,...,{({n^{13}})^m}\} $. Then we set
$$h({\mathbf{a}}) = {\mathbf{a}}\bmod {n^{13}} = ({a_1}\bmod {n^{13}},{a_2}\bmod {n^{13}},...,{a_m}\bmod {n^{13}})$$
Where ${\mathbf{a}} \in {\{ 0,...,M - 1\} ^m}$.
We replace the mapping $f$ in previous reduction by the mapping $g \circ h$. Next we will proof that the new mapping $g \circ h$ also meets the two conditions but makes the range small.
\\\indent Consider the special condition that for every ${\mathbf{b}},{\mathbf{b'}}$ that ${\bf{b}} - {\bf{b'}} = ({\alpha _1},...,{\alpha _m})$ and ${\mathbf{b}},{\mathbf{b'}} \in X$, there exists $g({\bf{b}}) - g({\bf{b'}}) = g(({\alpha _1},...,{\alpha _m}))$, as ${\mathbf{u}} = \sum\limits_{i = 1}^m {{\alpha _i}{{\mathbf{b}}_i}}  \in {\Lambda _q}({{\mathbf{A}}^T})$, then we get that ${\alpha _i} \le m{q^2}$ for $i \in \{ 1,...,m\} $ which implies $({\alpha _1},...,{\alpha _m}) \in X$, we can get $({\alpha _1},...,{\alpha _m})$ from $g(({\alpha _1},...,{\alpha _m}))$ because $g$ is bijective.
\\\indent Consider the general condition that for any vectors ${\mathbf{a}},{\mathbf{a'}}$ that satisfies ${\bf{a}} - {\bf{a'}} = ({\alpha _1},...,{\alpha _m})$, $h({\mathbf{a}}) = {\mathbf{b}}$ where ${\mathbf{b}} \in X$, the first entry of ${\mathbf{a}}$ is ${a_1} = {t_1}{n^{13}} + {b_1}$ where ${t_1} \in Z$ , as ${\mathbf{a}} - {\mathbf{a}}' = ({\alpha _1},...,{\alpha _m})$, we can get ${a_1}' = {a_1} - {\alpha _1}$. For ${b_1} \le {n^{13}}$, there exists ${b_1}'$ such that ${b_1} - {b_1}' = {\alpha _1}$ with possibility $1 - \frac{{m{q^2}}}{{{n^{13}}}} = 1 - \frac{{\log n}}{{{n^2}}}$, hence we get ${a_1}' = {t_1}{n^{13}} + {b_1}'$ and the first entry of $h({\mathbf{a}}')$ is ${b_1}'$. Hence for $\forall {\mathbf{a}},{\mathbf{a'}}$, there exists ${\mathbf{b}},{\mathbf{b'}} \in X$ that ${\bf{b}} - {\bf{b'}} = ({\alpha _1},...,{\alpha _m})$ with possibility ${(1 - \frac{{\log n}}{{{n^2}}})^m}$ such that $h({\mathbf{a}}) - h({\mathbf{a'}}) = {\mathbf{b}} - {\mathbf{b}}'$, this implies that $g \circ h({\mathbf{a}}) - g \circ h({\mathbf{a}}') = g(({\alpha _1},...,{\alpha _m}))$.
\\\indent Now consider the implementation of the mapping $g \circ h$, as $g$ is also a one-to-one mapping, for any ${\mathbf{a}} \in {\{ 0,...,M - 1\} ^m}$, it can be implemented as a unitary transform ${U_f}:\left| {\mathbf{a}} \right\rangle  \to \left| {f({\mathbf{a}})} \right\rangle $ as before, however $h({\mathbf{a}}) = ({a_1}\bmod {n^{13}},{a_2}\bmod {n^{13}},...,{a_m}\bmod {n^{13}})$ is not bijective, hence we have to add extra $B = m\left\lceil {13\log n} \right\rceil $ quantum qubits to construct a standard quantum oracle ${U_h}$ for $h$: $\left| {\mathbf{a}} \right\rangle \left| {\mathbf{0}} \right\rangle  \to \left| {\mathbf{a}} \right\rangle \left| {{\mathbf{0}} \oplus h({\mathbf{a}})} \right\rangle $.
\\\indent Here we give the whole procedure of solving LWE problem. Firstly, we get the quantum states $\left| {{\varphi _2}} \right\rangle  = \frac{1}{{\sqrt 2 }}(\left| {0,{\mathbf{a}}} \right\rangle  + \left| {1,{\mathbf{a}}'} \right\rangle )$ with ${\mathbf{a}} - {\mathbf{a}}' = ({\alpha _1},...,{\alpha _m})$ according to section 5. Then adding the quantum register and using the quantum oracle ${U_h}$, we can obtain $\left| {{\varphi _3}} \right\rangle  = \frac{1}{{\sqrt 2 }}(\left| 0 \right\rangle \left| {h({\mathbf{a}})} \right\rangle  + \left| 1 \right\rangle \left| {h({\mathbf{a}}')} \right\rangle )$ (qubits for $\left| {\mathbf{a}} \right\rangle $ are discarded). Finally, using the quantum oracle ${U_g}$, we can get $\left| {{\varphi _4}} \right\rangle  = \frac{1}{{\sqrt 2 }}(\left| 0 \right\rangle \left| {g \circ h({\mathbf{a}})} \right\rangle  + \left| 1 \right\rangle \left| {g \circ h({\mathbf{a}}')} \right\rangle )$, which is  the input of DCP over ${D_{{n^{13n\log n}}}}$. The following is the quantum circuit for the above procedure:
\begin{figure}[htbp]
\centering
\includegraphics[width=10.66cm,height=3.83cm]{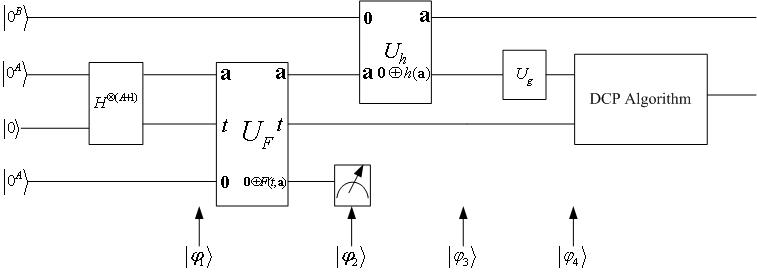}
\caption{the quantum circuit for solving LWE problem}
\label{fig2}
\end{figure}
\section{Conclusions}
In this work we give a reduction from LWE problem to the dihedral coset problem. First we obtain the property of the solution to LWE problem based on the error vector ${\mathbf{e}}$ from the discrete Gaussian distribution; then we present a quantum algorithm to get an input to the two point problem. Iterating this algorithm polynomial times gives the complete input of the two point problem; finally we present a new reduction from two point problem to dihedral coset problem which bringing the size of dihedral group down from ${D_{{2^{(n + 1)n}}}}$ to ${D_{{n^{13n\log n}}}}$.



\end{document}